\documentstyle[prc,aps,manuscript,epsf]{revtex}

\begin{document}

\draft

\title{COLLISIONAL RELAXATION OF COLLECTIVE MOTION
IN A FINITE FERMI LIQUID}

\author{ V.M.Kolomietz$^{1)}$, S.V.Lukyanov$^{1)}$, 
V.A.Plujko$^{1)}$ and S.Shlomo$^{2)}$}

\address{$^{1)}$Institute for Nuclear Research,
Prosp. Nauki 47, 252028 Kiev, Ukraine}
\address{$^{2)}$Cyclotron Institute, Texas A\&M University,
College Station, Texas 77843}


\maketitle

\begin{abstract}
Finite size effects in the equilibrium phase space density distribution 
function are taken into account for alculations of the 
relaxation of collective motion in finite nuclei. Memory effects
in the collision integral and the diffusivity and the quantum 
oscillations of the equilibrium distribution function in momentum 
space are considered. It is shown that a smooth diffuse (Fermi-type)
equilibrium distribution function leads to a spurious contribution to the 
relaxation time.  The residual quantum oscillations of the equilibrium
distribution function eliminates the spurious contribution. It ensures 
the disappearance of the gain and loss terms in the collision integral 
in the ground state of the system and strongly reduces the internal 
collisional width of the isoscalar giant quadrupole resonances.
\end{abstract}

\vskip 1 cm

\pacs{PACS number(s): 21.60.Ev, 24.30.Cz}

\section{Introduction}
The relaxation of nuclear collective motion toward thermal equilibrium
have been described in great detail within the framework of the
kinetic theory, taking into account the collision integral 
\cite{Be.ZP.78,NiSi,Kob,To2,KoMgPl,KoPlSh1,SaYo,KoPlSh2,KoLuPl}. 
In this theory, the damping of the collective motion appears
due to the interparticle collisions on the dynamically deformed
Fermi surface. It has already been shown by Landau \cite{land1,LiPi} that both
temperature and memory effects are extremely important for successful
applications of kinetic theory to the relaxation processes in a
Fermi system. However, only very little attention has been paid to the 
study of the peculiarities of the collision integral in a
finite Fermi system  caused by particle reflections
on the boundary.

In kinetic theory, the collision integral depends crucially on the
phase-space distribution function $f(\vec{r},\vec{p})$. The main
aim of this paper is to apply the quantum Wigner phase-space distribution
function, also known as the Wigner transform \cite{Wi.PR.32}, to the
evaluation of the collision integral in a finite Fermi system. The
Wigner distribution function (WDF) is defined as the Fourier transform
of the one-body density matrix over the relative coordinates. It possesses
several nice properties \cite{Sh.NC.85,CaZa.RMP.83} which justify its
interpretation as the quantum mechanical analog of the classical phase-space
distribution function. The WDF is useful in providing a reformulation
of quantum machanics in terms of classical concepts \cite{Sh.PL.82,Sh.JP.84}
and a good starting point for semiclassical approximations \cite{RiSc.b.80}.

Traditionally, the equilibrium phase space density distribution function 
$f_{eq}(\vec{r}, \vec{p})$ in the collision integral is replaced by
the one, $f_{eq,TF}(\vec{r},\vec{p})$, taken in Thomas-Fermi approximation
\begin{equation}
f_{eq,TF}(\vec{r},\vec{p})= \theta(\lambda-E(\vec{r}, \vec{p})),   
\label{ftf}
\end{equation}
where $\theta (x)$ is the step function and $E(\vec{r}, \vec{p})$ is the 
classical single-particle energy. This is reasonable for an infinite Fermi 
system. In the case of a finite Fermi system, the quantum distribution 
function $f_{eq}(\vec{r},\vec{p})$ fluctuates strongly and contains the 
diffusivity of the Fermi surface even at zero temperature  
\cite{ShPr.NP.81,PrShKo,PrShNiBoSe.NP.82,ZvSa,DuRaSc,KoPl377}. 
Both these features are due to particle reflections on potential walls.

The diffusivity of the Fermi surface in momentum space increases
in the vicinity of the nuclear surface \cite{DuRaSc} and thus enhances 
the effective particle scattering there because of the decrease of the
Pauli blocking effect. It can be shown \cite{KoLuPl} that
the use of a simple Fermi distribution function 
\begin{equation}
f_{eq,F}(\vec{r}, \vec{p})=\left(1+\exp\left[{E(\vec{r}, \vec{p}) - 
   \lambda \over a(r)}\right]\right)^{-1},
\label{ff}
\end{equation}
with $r$-dependent diffusivity parameter $a(r)$, instead of the
$\theta$-function of Eq. (\ref{ftf}), for the equilibrium distribution 
function $f_{eq}(\vec{r}, \vec{p})$ in the collision integral, 
leads to a significant enhancement of the damping of the nuclear giant 
multipole resonances. However, there is a conceptual
disadvantage for the application of the Fermi distribution function
$f_{eq,F}(\vec{r}, \vec{p})$ with $a(r) \neq 0$ to the collision
integral. Namely, with this function the gain and loss terms in the 
collision integral are each nonzero for the ground state of the
system, where the probability current should be absent by definition. 
We show in this work that in order to overcome this difficulty the 
smooth quantum distribution function $\tilde{f}_{eq}(\vec{r}, \vec{p})$
should be used for $f_{eq}(\vec{r}, \vec{p})$ in the collision
integral in the kinetic Landau-Vlasov equation. In contrast to the
Fermi distribution function $f_{eq,F}(\vec{r}, \vec{p})$, the smooth 
quantum distribution function $\tilde{f}_{eq}(\vec{r}, \vec{p})$
contains the residual oscillations \cite{PrShKo,KoPl377} ensuring
the above-mentioned condition for the disappearance of the gain and
loss terms in the collision integral for the ground state and reducing 
the internal collisional width of the giant multipole resonances.

In this paper we pay attention mainly to finite size and memory
effects in the  relaxation processes in finite Fermi systems.
In Sec.II we obtain a general expression for the width of the
giant multipole resonances at zero temperature of the nucleus
starting from the collisional Landau-Vlasov equation.
In Sec. III we study the influence of the memory effects and the
diffusivity of the Fermi surface on the relaxation time. We use
the smeared-out Wigner distribution function
$\tilde{f}_{eq}(\vec{r}, \vec{p})$ of the three-dimensional 
harmonic oscillator and of the Woods-Saxon potential for the calculations 
of the collision integral. A summary and conclusions are given in Sec. IV.

\bigskip

\section{Damping of collective excitations in kinetic theory}

The kinetic equation for a small variation $\delta f (\vec{r}, \vec{p}, t)$ 
of the distribution function can be transformed to a set (infinite) of 
equations for the moments of $\delta f (\vec{r}, \vec{p}, t)$ in 
${\vec p}$ space, namely, the local particle density $\delta \rho$, velocity
field $\vec{u}$, pressure tensor $\pi_{\alpha \beta}$, etc.,
see \cite{LiPi,AbKh}. The first-order moment of the kinetic equation 
has the form of the Euler-Navier-Stokes equation and is given by
\cite{AbKh,KoTa}
\begin{equation}
m \rho_{eq} {\partial \over\partial t} u_{\alpha} +
\rho_{eq} { \partial \over\partial r_{\nu}}
\left({\delta^{2} {\cal E} \over \delta \rho^{2}}\right)_{eq}
\delta\rho + { \partial \over\partial r_{\nu}}\pi_{\nu\alpha} = 0,
\label{eiler}.
\end{equation}
Here and in the following expressions, repeated greek indices 
$\alpha,\beta,\nu=1,2,3$ are to be understood as summed over.
The variation of the local particle density $\rho$ and the
velocity field $u_\alpha$ in Eq. (\ref{eiler}) are defined by
\begin{equation}
\delta\rho=g\displaystyle \int {d\vec{p} \over (2\pi\hbar)^3}\delta f,
\,\,\,\,\,\,\,
\vec{u}=g \displaystyle \int {d\vec{p} \over (2\pi\hbar)^3}
     {\vec{p}\over m}\delta f.
\label{u}
\end{equation}
The quantity $\pi_{\alpha\beta}$ is the deviation of the pressure tensor 
from its equilibrium part $P_{eq}$ due to the Fermi-surface deformation 
\begin{equation}
\pi_{\alpha\beta}= {g \over m}\displaystyle \int {d\vec{p} \over (2\pi\hbar)
   ^3}(p_{\alpha}-mu_{\alpha})(p_{\beta}-mu_{\beta})\delta f
\label{piab}
\end{equation}
and the equilibrium pressure, $P_{eq}$, is given by
\begin{equation}
P_{eq} = g \displaystyle \int {d\vec{p} \over (2\pi\hbar)^3}
           {p^{2}\over 2m} f_{eq} 
           \equiv {2 \over 3} {\cal E}_{kin},
\label{p}
\end{equation}
where ${\cal E}_{kin}$ is the kinetic energy density.
The internal energy density ${\cal E}$ in Eq. (\ref{eiler}) contains
both kinetic and potential energy densities: 
$ {\cal E}={\cal E}_{kin}+{\cal E}_{pot}$,
where ${\cal E}_{pot}$ is the potential energy density.

Eq. (\ref{eiler}) is not closed because it contains the pressure
tensor $\pi_{\alpha \beta}$ given by the second-order moment of 
the distribution function $\delta f (\vec{r}, \vec{p}, t)$, Eq.
(\ref{piab}).
We will follow the nuclear fluid dynamic approach of Refs. 
\cite{NiSi,Kob,KoPlSh1} and take into account the dynamic 
Fermi-surface distortions up to multipolarity $l = 2$. The second 
$\vec{p}$ moment of the kinetic equation leads then to a closed 
differential equation for the pressure tensor $\pi_{\alpha \beta}$.
Namely  (see Refs. \cite{Kob,KoPlSh1,KoPlSh2,KoLuPl,KiKoSh}),
\begin{equation}
{\partial \over\partial t}\pi_{\alpha\beta} +
P_{eq} \Big({\partial u_{\alpha}\over \partial r_{\beta}}+
{\partial u_{\beta}\over \partial r_{\alpha}}-
{2 \over 3}\delta_{\alpha\beta}{\partial u_{\alpha}\over \partial
r_{\beta}}\Big) = -{\pi_{\alpha\beta} \over \tau_{2}}.
\label{tre}
\end{equation}
The local relaxation time $\tau_2$ in Eq. (\ref{tre}) is caused by
the interparticle scattering on the deformed Fermi surface: 
\begin{equation}
{1 \over \tau_{2}}= -{\displaystyle \int d\vec{p} p^2 Y_{20} \
\delta St \over \displaystyle \int d\vec{p} p^2 Y_{20}  \ \delta f}.
\label{tau2}
\end{equation} 
Here $\delta St \equiv \delta St (\vec{r},\vec{p},t)$
is a collision integral linearized in $\delta f$. 
In the case of small eigenvibrations with eigenfrequency 
$\omega=\omega_{0}+i\Gamma/2\hbar$, where 
$\omega_0$ and $\Gamma$ are real, the collision integral can be
transformed, taking into account also memory effects, as 
\cite{KoMgPl,KoPl992}
\begin{eqnarray}
\delta St (\vec{r},\vec{p},t)=
\displaystyle \int {g\,d\vec{p}_{2} d\vec{p}_{3} 
d\vec{p}_{4} \over (2 \pi \hbar)^{6}}  \nonumber \\
\times W(\{\vec{p}_{j}\}) \sum_{j=1}^{4}{\delta Q \over \delta f_{j}}
\Big\vert_{eq}
\delta f_{j} {1 \over 2} \left(\delta (\Delta E + \hbar\omega_{0} )+ \delta 
(\Delta E - \hbar\omega_{0} ) \right) \delta (\Delta \vec{p}),
\label{dst}
\end{eqnarray}
where $W(\{\vec{p}_{j}\})$ 
is the probability of the scattering of nucleons near the 
Fermi surface, $g = 4$ is the spin-isospin degeneracy factor and
$$Q = (1-f_{1})(1-f_{2})f_{3}f_{4} - f_{1}f_{2}(1-f_{3})(1-f_{4});$$
$$\Delta E = E_{1}+E_{2}-E_{3}-E_{4}; \ \ \ \ \ \ \ \ \Delta \vec{p}=
\vec{p}_{1}+ \vec{p}_{2}-\vec{p}_{3}-\vec{p}_{4};$$
$E_{j}=p^{2}_{j}/2m+V(\vec{r}_j)$ is the classical single-particle energy,
$\vec{p}_{1} \equiv \vec{p}$ and $V(\vec{r})$ is the nuclear
mean field.

We will follow the arguments of the Fermi-liquid theory of Ref. 
\cite{AbKh} and assume that the dynamical component of the distribution 
function $\delta f (\vec{r}, \vec{p}, t)$ has the form
\begin{equation}
\delta f(\vec{r},\vec{p},t) = -{\partial f_{eq}\over \partial E}
\nu (\vec{r},\vec{p},t),
\label{df}
\end{equation}
where $\nu ( \vec{r}, \vec{p}, t)$ are unknown functions.
The functions $\nu ( \vec{r}, \vec{p}, t)$ depend on the 
orientation $\hat{p}$ only because of the sharp energy dependence of 
the factor ${\partial f_{eq}/\partial E}$ in Eq. (\ref{df}), which is
localized at the Fermi momentum $p_F(r)$. We point out that the smooth 
region of the equilibrium distribution function $f_{eq}(\vec{r},
\vec{p})$ in momentum space appears in a classical forbidden region
at $E < V(\vec{r})$. However, this region is absent in the space
integral in Eq. (\ref{dst}). Thus,
\begin{equation}
\nu ( \vec{r}, \vec{p}, t) \approx \nu (\vec{r}, p_F(r), \hat{p}, t)
=\sum^{}_{lm}\nu _{lm}(\vec{r},t)Y_{lm}(\hat{p}).
\label{nu}
\end{equation}

An exact evaluation of the nine-dimension integral in 
$\vec{p}$ space in Eq. (\ref{dst}) is a very complicated problem. 
We will follow the Abrikosov-Khalatnikov method \cite{AbKh} improved 
in Ref. \cite{VOG}. Let us assume that the scattering probability
$W(\{\vec{p}_{j}\})$ in Eq. (\ref{dst}) depends on two scattering
angles $\theta$ and $\phi$ only, (see Ref. \cite{AbKh}), where
$\theta$ is the angle between $\vec{p}_{1}$ and $\vec{p}_{2}$, 
and $\phi $ is the angle between the planes formed by
($\vec{p}_{1},\vec{p}_{2}$)  and  ($\vec{p}_{3},\vec{p}_{4}$), i.e.,
\begin{equation}
W(\{\vec{p}_{j}\}) \approx W(\{p_j = p_F (r),\,\theta,\,\phi\}).
\label{w1}
\end{equation}
To evaluate the collision integral Eq. (\ref{dst}), we will use 
the Abrikosov-Khalatnikov transformation \cite{AbKh,VOG}:
\begin{equation}
\int {d\vec{p}_{2} d\vec{p}_{3} d\vec{p}_{4}\over (2\pi\hbar)^6}\,
\delta (\Delta \vec{p})\, (...) \approx
{m^3\over 2\,(2\pi\hbar)^6}
\int {d\Omega d\phi_{2}
\over \cos \theta/2} dE_2\,dE_3\,dE_4\, (...),
\label{ach}
\end{equation}
where $d\Omega=\sin \theta d\theta d\phi$ and
$\phi_2$ is the azimuthal angle of the momentum $\vec{p}_2$ 
in the coordinate system with the $z$ axes along $\vec{p}_{1}$.
We point out that the angle $\phi$ varies only from $0$ to $\pi$ because
the particles are indistinguishable. 

Using Eq. (\ref{ach}), the collision integral 
Eq. (\ref{dst}) can be written in the form
\begin{eqnarray}
\delta St (\vec{r},\vec{p},t)
= - {m^3 \over 16 \pi^{4} \hbar^{6}} \nonumber \\
\times \sum_{l,m} \nu_{lm}(\vec{r},t) Y_{lm} (\hat{p})
\sum_{j=1}^{4} \langle W(\theta, \phi) P_{l}(\cos \theta_{j})\rangle 
\left(I_{j}^{(+)} + I_{j}^{(-)}\right).
\label{dSt2}
\end{eqnarray}
The symbol $\langle ...\rangle $ denotes the averaging
over angles of the relative momentum of the colliding particles
\begin{equation}
\langle W(\theta , \phi) P_{l}(\cos \theta_j) \rangle =
\displaystyle \int^{\pi }_{0} d\theta \frac{\sin \theta}{\cos \theta/2}
\int^{\pi }_{0} \frac{d\phi}{2\pi} W(\theta ,\phi ) P_{l}(\cos \theta_j),
\label{scob}
\end{equation}
where
$\cos \theta _{j}\equiv (\hat{p}_{j}\hat{p}_{1})$, \,\,\,\,
{\rm i.e.},\,\,\,\,
$\theta _{2}\equiv \theta$, 
and
$$ 
\cos \theta _{3} =
(\cos (\theta/2))^2 + (\sin (\theta/2))^2 \cos \phi , \ \ \
\cos \theta _{4} =
(\cos (\theta/2))^2 - (\sin (\theta/2))^2 \cos \phi ,
$$
$P_{l}(\cos \theta)$ is a Legendre polynomial and
$$I_{j}^{(\pm)}=\displaystyle \int_{V_{eq}}^{\infty} dE_{2} dE_{3} dE_{4}
{\partial f_{eq,j} \over \partial E_{j}} {\delta Q \over 
\delta f_{j}}\Big\vert_{eq} \delta(\Delta E \pm \hbar\omega_{0}). $$

We now return to the dynamical equation (\ref{eiler}).
Let us introduce the displacement field $\vec{\chi}(\vec{r},t)$
related to the velocity field $\vec{u}(\vec{r},t)$ as
\begin{equation}
\vec{u}(\vec{r},t)
={\partial \over \partial t} \vec{\chi}(\vec{r},t).
\label{pu}
\end{equation}
We look for the displacement field $\vec{\chi}(\vec{r},t)$
in the following separable form:
\begin{equation}
\vec{\chi}(\vec{r},t) = \beta(t) \vec{v}(\vec{r}),
\label{v}
\end{equation}
where $\beta (t) = \beta_0 e^{i\omega t}$.
Substituting Eqs. (\ref{pu}) and (\ref{v}) into Eq. (\ref{eiler}),
one obtains the equation of motion for the collective variable
$\beta (t)$:
\begin{eqnarray}
m\rho_{eq}v_{\alpha}\ddot{\beta}+{1\over\omega}{\partial \over\partial
  r_{\nu}}\left({\rm Im} 
  \Pi_{\nu\alpha}^{\omega}\right) \dot{\beta}+
{\partial \over \partial r_{\nu}}\left({\rm Re}
\Pi_{\nu\alpha}^{\omega}\right)\beta- 
   \nonumber \\
-\rho_{eq}{\partial \over \partial r_{\alpha}}\left[\left({\delta^{2}
   {\cal E} \over \delta \rho^{2}}\right)_{eq}{\partial \over \partial
    r_{\nu}}(\rho_{eq}v_{\nu})\right]\beta=0,
\label{urdv2}
\end{eqnarray}
where we have used the following form for the  traceless part 
$\pi_{\alpha\beta}(\vec{r},t)$ of the momentum flux tensor
\begin{equation}
\pi_{\alpha\beta}(\vec{r},t) \equiv \Pi_{\alpha\beta}^{\omega}(\vec{r})
\beta_0 e^{i\omega t} =
\beta(t) {\rm Re}\Pi_{\alpha\beta}^{\omega}(\vec{r})+
{1 \over \omega} \dot{\beta}(t) {\rm Im}
\Pi_{\alpha\beta}^{\omega}(\vec{r}).
\label{piri}
\end{equation}
Multiplying Eq. (\ref{urdv2}) by $v_\alpha$, summing over $\alpha$
and integrating over $\vec{r}$ space, we obtain the dispersion
equation for the eigenfrequency $\omega$:
\begin{eqnarray}
-B\omega^{2}+i\omega A(\omega)+\widetilde{C}(\omega)=0.
\label{urdv4}
\end{eqnarray}
Here, $B$ is the hydrodynamical mass coefficient \cite{BoMo} with
respect to the collective variable $\beta (t)$:
\begin{equation}
B=\displaystyle  m\int d\vec{r} \rho_{eq} v^{2}.
\label{mass}
\end{equation}
The dissipative term $A(\omega)$ and the stiffness coefficient
$\widetilde{C}(\omega)=C+C^{\prime}(\omega)$ are given by
\begin{equation}
A(\omega)={1 \over \omega} \displaystyle \int d\vec{r} v_{\alpha}{\partial
  \over\partial r_{\nu}}\left({\rm Im} \Pi_{\nu\alpha}^{\omega}\right) 
\label{A}
\end{equation}     
and
\begin{eqnarray}
C= \displaystyle \int d\vec{r} 
 \left({\delta^{2} {\cal E} \over \delta \rho^{2}}
  \right)_{eq}\left[{\partial \over \partial r_{\nu}}(\rho_{eq}v_{\nu})
  \right]^{2},
\label{CLD}\\
C^{\prime}(\omega)=\displaystyle \int d\vec{r} v_{\alpha}{\partial \over
   \partial r_{\nu}}\left({\rm Re} \Pi_{\nu\alpha}^{\omega}\right).
\label{CF}
\end{eqnarray}
We point out that the definition of the stiffness coefficient $C$ 
coincides with the one
for the stiffness coefficient in the traditional liquid drop model (LDM)
for the nucleus. In contrast, the additional contribution
$C^{\prime}(\omega)$ to the stiffness coefficient is absent in the LDM and
represents the influence of the dynamical Fermi-surface distortion
on the conservative forces in the Fermi system. Finally, the dissipative
term $A(\omega)$ appears due to the interparticle scattering on the
distorted Fermi surface.

To evaluate the coefficients $A(\omega)$ and $C^{\prime}(\omega)$,
we will use the third equation of motion (\ref{tre}). Let us
rewrite Eq. (\ref{tre}) in the form
\begin{equation}
{\partial \over \partial t} \pi_{\alpha\beta}+
{1 \over \tau_{2}}\pi_{\alpha\beta}=-P_{eq}\Lambda_{\alpha\beta} ,
\label{ddd}
\end{equation}
where
\begin{equation}
\Lambda_{\alpha\beta}={\partial u_{\alpha}\over \partial r_{\beta}}+
{\partial u_{\beta}\over \partial r_{\alpha}}-
{2 \over 3}\delta_{\alpha\beta}{\partial u_{\alpha}\over \partial
r_{\beta}}.
\label{lamd}
\end{equation}
Taking Eqs. (\ref{pu}), (\ref{piri}) and (\ref{ddd}), one obtains
\begin{equation}
\dot{\beta}{\rm Re}\Pi^{\omega}_{\alpha\beta}+
{1 \over \omega}\ddot{\beta}{\rm Im}
\Pi^{\omega}_{\alpha\beta}+{1 \over \tau_{2}}
\beta {\rm Re} 
\Pi^{\omega}_{\alpha\beta}+{1 \over \omega\tau_{2}}\dot{\beta}
{\rm Im}
\Pi^{\omega}_{\alpha\beta}=-P_{eq}\dot{\beta}\overline{\Lambda}_{\alpha\beta}.
\label{urd}
\end{equation}
Here,
\begin{equation}
\overline{\Lambda}_{\alpha\beta}
={\partial v_{\alpha}\over \partial r_{\beta}}+
{\partial v_{\beta}\over \partial r_{\alpha}}-
{2 \over 3}\delta_{\alpha\beta}{\partial v_{\alpha}\over \partial r_{\beta}}.
\label{lamd2}
\end{equation}
Eq. (\ref{urd}) can be represented as a set of equations of
motion for both the real and the imaginary parts of $\omega$:
\begin{eqnarray}
- \Gamma Re\Pi^{\omega}_{\alpha\beta} - 
2\hbar \omega_{0} Im\Pi^{\omega}_{\alpha\beta}
  + {2\hbar \over \tau_{2}} Re\Pi^{\omega}_{\alpha\beta} - \Gamma P_{eq}
 \overline{\Lambda}_{\alpha\beta}=0,  \\
\label{re}
 2\hbar \omega_{0} Re\Pi^{\omega}_{\alpha\beta} - 
  \Gamma Im\Pi^{\omega}_{\alpha\beta}+
  {2\hbar \over \tau_{2}}Im\Pi^{\omega}_{\alpha\beta} + 
  2\hbar \omega_{0}P_{eq}
  \overline{\Lambda}_{\alpha\beta}=0,
\label{im}
\end{eqnarray}
where ${\rm Im} (\omega)\equiv \Gamma /(2\hbar)$.\\

In the case of small damped collective motion, we find from
Eq. (\ref{urd})
\begin{eqnarray}
{\rm Re}
\Pi^{\omega}_{\alpha\beta}=-P_{eq}{(\omega_{0}\tau_{2})^{2} \over 1+
 (\omega_{0}\tau_{2})^{2}}\overline{\Lambda}_{\alpha\beta},
\label{repi} \\
{\rm Im}\Pi^{\omega}_{\alpha\beta}=-P_{eq}{\omega_{0}\tau_{2} \over 1+
(\omega_{0}\tau_{2})^{2}}\overline{\Lambda}_{\alpha\beta}.
\label{impi}
\end{eqnarray}
Finally, we have from Eqs. (\ref{A}), (\ref{CF}), (\ref{repi}) and
(\ref{impi}), (see also Ref. \cite{KiKoSh}),
\begin{eqnarray}
A(\omega_{0})=\displaystyle \int d\vec{r}P_{eq}{\tau_{2} \over
 1+(\omega_{0} \tau_{2})^{2}}\overline{\Lambda}_{\alpha\nu}
  {\partial v_{\alpha} \over\partial r_{\nu}}, 
\label{aa}  \\
C^{\prime}(\omega_{0})=\displaystyle \int d\vec{r} P_{eq}{(\omega_{0}
 \tau_{2})^2 \over 1+(\omega_{0}\tau_{2})^{2}} \overline{\Lambda}_{\alpha\nu}
 {\partial v_{\alpha} \over\partial r_{\nu}}.
\label{cc}
\end{eqnarray}

In the same case of small damped collective motion, the dispersion 
equation (\ref{urdv4}) is transformed as
\begin{equation}
-B(\omega^{2}_{0} + i {\Gamma \over \hbar}\omega_{0})+ 
i \omega_{0} A(\omega_{0}) +  \widetilde{C}(\omega_{0}) = 0.
\label{osc}
\end{equation}
Thus,
\begin{eqnarray}
\omega^{2}_{0}={\widetilde{C}(\omega_{0}) \over B} \ \ , \ \
\Gamma=\hbar {A(\omega_{0}) \over B} .
\label{gamv}
\end{eqnarray}
Using Eqs. (\ref{gamv}), (\ref{aa}) and (\ref{cc}) we  obtain
the  width $\Gamma$ as
$$
\Gamma=\hbar \omega_{0} {\omega_{0} A(\omega_{0}) \over
 \widetilde{C}(\omega_{0})}
$$
\begin{equation}
 = \hbar \omega_{0} { \displaystyle \int
d\vec{r} P_{eq}\overline{\Lambda}_{\alpha\nu}
  (\partial v_{\alpha} /\partial r_{\nu}) 
  \omega_{0} \tau_{2} /[1+(\omega_{0} \tau_{2})^{2}]
  \over C +
 \displaystyle \int d\vec{r} P_{eq}\overline{\Lambda}_{\alpha\nu}
 (\partial v_{\alpha} /\partial r_{\nu})(\omega_{0} \tau_{2})^2 /
 [ 1+(\omega_{0}\tau_{2})^{2}]}.
\label{gam1}
\end{equation}
     
In the case of an isoscalar giant quadrupole resonance (GQR)
one can assume \cite{Kob,To2,RiSc.b.80} that the displacement field 
$\vec{v}(\vec{r})$ is the irrotational one and is given by
$\vec{v}=(-x,-y,2z)$. We have then $\displaystyle \overline
{\Lambda}_{\alpha\nu} {\partial v_{\alpha}\over\partial r_{\nu}} = 8$.
Furthermore, the LDM stiffness coefficient $C$ gives a negligible 
contribution to the total value $\widetilde{C}(\omega)$ \cite{Kob,RiSc.b.80} 
and Eq. (\ref{gam1}) is transformed as
\begin{equation}
\Gamma \cong \hbar\omega_{0} {\displaystyle\int d\vec{r}P_{eq}
\omega_{0}\tau_{2}/[1+(\omega_{0}\tau_{2})^{2}] \over 
\displaystyle \int d\vec{r}P_{eq}
 (\omega_{0}\tau_{2})^{2}/ [1+(\omega_{0}\tau_{2})^{2}]}.
\label{gam2}
\end{equation}
Finally, in the rare collision regime ($\omega_{0} \tau \gg 1$) 
Eq. (\ref{gam2}) is reduced as
\begin{equation}
\Gamma_r \simeq \hbar\displaystyle \int d\vec{r} {P_{eq}
  \over \tau_{2}} / \displaystyle \int d\vec{r} P_{eq} .
\label{gamsr}
\end{equation}

\section{Results of numerical calculations}

To study the space distribution of the collective damping,
we will introduce the local damping parameter $\xi (r, \omega_0)$ 
related to the width $\Gamma$:
\begin{equation}
\Gamma \equiv \displaystyle \int d\vec{r} \xi (r,\omega_{0}),
\label{loks}
\end{equation}
where (see Eq. (\ref{gam2}))
\begin{equation}
\xi (r,\omega_{0}) = {\hbar\omega_{0} P_{eq}
\omega_{0}\tau_{2}/[1+(\omega_{0}\tau_{2})^{2}] \over 
\displaystyle \int d\vec{r}P_{eq}
 (\omega_{0}\tau_{2})^{2}/ [1+(\omega_{0}\tau_{2})^{2}]},
\label{xi}
\end{equation}
is always a positive quantity.
The local relaxation time 
$\tau_2 \equiv \tau_{2}(r,\omega_{0})$ can be obtained by
substituting  Eq. (\ref{dSt2})  into Eq. (\ref{tau2}) and is given by
\begin{equation}
{1 \over \tau_{2}(r,\omega_{0})} = -{m^3 \over 16\pi^{4}\hbar^{6}}
\langle W(\theta , \phi) \rangle {R^{(+)}+R^{(-)} \over \displaystyle
\int^{\infty}_{V_{eq}} dE 
(E-V_{eq})^{3/2}{\partial f_{eq} \over \partial E}} \equiv 
\gamma (r,\omega_0), 
\label{tau22}
\end{equation}
where
\begin{eqnarray}
R^{(\pm)}=\displaystyle \int^{\infty}_{V_{eq}} 
dE_{1} dE_{2} dE_{3} dE_{4}
(E_{1}-V_{eq})^{3/2} \delta(\Delta E \pm \hbar\omega_{0}) 
\nonumber \\
\times \left({\partial f_{eq, 1} \over \partial E_{1}}
 {\delta Q \over \delta f_{eq, 1}}+  c_{2}{\partial f_{eq, 2} \over 
\partial E_{2}}
 {\delta Q \over \delta f_{eq, 2}} + (1+c_{2}-d_{2}){\partial f_{eq, 3} 
 \over \partial E_{3}} {\delta Q \over \delta f_{eq, 3}} \right),
\label{Rpm}
\end{eqnarray}
\smallskip
and the coefficients $c_{2}$ and $d_{2}$  are given by
\begin{equation}
c_{2}=\langle WP_{2}(\cos \theta )\rangle /\langle W\rangle ,
\ \ \ \ 
d_{2}=\langle 3\,W\,\sin^4{\theta\over 2}\,\sin^2 \phi\rangle /
\langle W\rangle. 
\label{cd}
\end{equation} 

To evaluate the relaxation time (\ref{tau22}), we will study, first
of all, the collision integral $St_{eq} (\vec{r},\vec{p})$
in the ground state of the system which is given by (see also Eq. 
(\ref{dst}))
\begin{eqnarray}
St_{eq}(\vec{r},\vec{p})=\displaystyle \int {g\,d\vec{p}_{2} d\vec{p}_{3} 
d\vec{p}_{4} \over (2 \pi \hbar)^{6}} \,
 W(\{\vec{p}_{j}\}) \,Q\Big\vert_{eq}
\delta (\Delta E )\, \delta (\Delta \vec{p}).
\label{st}
\end{eqnarray}
We will introduce also the total gain and loss fluxes of the 
probability in the ground state. They are given, respectively, by
\begin{equation}
J_{eq, gain} = \displaystyle \int d\vec{r}_{1}  
{g\,d\vec{p}_{1} d\vec{p}_{2} d\vec{p}_{3} d\vec{p}_{4} \over
(2\pi\hbar)^9}\,
W(\{\vec{p}_{j}\}) \ [1-f_{eq, 1}][1 - f_{eq, 2}] f_{eq,
3}f_{eq, 4}\, \delta (\Delta E ) \, \delta (\Delta \vec{p}),
\label{gain} 
\end{equation}
and
\begin{equation}
J_{eq, loss} = \displaystyle \int d\vec{r}_{1}  
{g\,d\vec{p}_{1} d\vec{p}_{2} d\vec{p}_{3} d\vec{p}_{4} \over
(2\pi\hbar)^9}\,
W(\{\vec{p}_{j}\}) \ [1-f_{eq, 3}][1 - f_{eq, 4}] f_{eq,
1}f_{eq, 2}\, \delta (\Delta E ) \, \delta (\Delta \vec{p}).
\label{loss} 
\end{equation}
One can see from both definitions, Eqs. (\ref{gain}) and (\ref{loss}), that
$J_{eq, gain} = J_{eq, loss}$ and $St_{eq}(\vec{r},\vec{p}) = 0$,
as it should be for the equilibrium state of system.
Moreover, in the case of the ground state of the system, both fluxes
$J_{eq, gain}$ and  $J_{eq, loss}$ have to disappear separately.
This is not the case, however, if the Fermi distribution function
$f_{eq,F}$ (see Eq. (\ref{ff})) is used for the equilibrium
distribution function $f_{eq}$ of the ground state of the finite
Fermi system in Eqs. (\ref{gain}) and (\ref{loss}).
To avoid this disadvantage we will use the smeared-out
quantum distribution function $\widetilde{f}_{eq}$ in both
Eqs. (\ref{gain}) and (\ref{loss}). To control the disappearance
of the gain and loss fluxes, we will introduce the
relative contribution $q$ of the probability fluxes
$J_{eq, gain}$ or  $J_{eq, loss}$ evaluated with 
$f_{eq} = \widetilde{f}_{eq}$ to the corresponding values
evaluated with $f_{eq} =  f_{eq,F}$, i.e.,
\begin{equation}
q=J_{eq, gain}(\{\widetilde{f}_{eq}\})/J_{eq, gain}(\{f_{eq,F}\}) 
\equiv J_{eq, loss}(\{\widetilde{f}_{eq}\})/J_{eq, loss}(\{f_{eq,F}\}).
\label{q}
\end{equation} 

Below we will apply our approach to the isoscalar GQR. 
We will assume that the scattering probability $W$ 
in Eqs. (\ref{tau22}) and (\ref{cd}) is angle independent, i.e.,
$d_{2}=4/5$ and  $c_{2}=1/5$, and the magnitude of $W$ can be
obtained from the nuclear matter estimate of the  parameter
$\alpha \equiv  15\, \pi^{2} \,\hbar^{5}/m^3\, \langle W\rangle =
9.2 \,MeV$ from Ref. \cite{KoPlSh2}. We will also use the following
expression for the energy $\hbar\omega_0$ of the isoscalar GQR,
$\hbar\omega_{0}=60 A^{-1/3}$ MeV. The numerical calculations will 
be performed for both the spherical harmonic oscillator (HO) 
potential well and the Woods-Saxon (WS) potential.

\medskip
{\bf A. Spherical harmonic oscillator potential}
\medskip

We will use the harmonic oscillator potential in the form
\begin{equation}
V_{eq}(r)={1 \over 2}m \Omega^{2} r^{2} ,
\label{HO}
\end{equation}
where $\hbar\Omega \simeq 41A^{-1/3}$ MeV.
For "magic" nuclei in the absence of the spin-orbit interaction,
the smeared-out quantum distribution function
$\tilde{f}_{eq}(\vec{r}, \vec{p})$ in a HO potential
is given by \cite{PrShKo}
\begin{equation}
\widetilde{f}_{eq}(\vec{r},\vec{p}) \equiv \widetilde{f}_{eq}(\epsilon)=
8e^{-\displaystyle\epsilon} \sum_{k=0}^{\infty}(-1)^{k}L_{k}^{2}
(2\epsilon)\widetilde{n}_{k}.
\label{fw}
\end{equation}
Here, $L^{n}_{k}(\epsilon)$ is the associated Laguerre polynomial and
$\epsilon=p^{2}/m\hbar\Omega+m\Omega r^{2}/\hbar\equiv 
2E/\hbar\Omega$ is the dimensionless energy parameter. The smooth
occupation numbers $\widetilde{n}_{k}$ are introduced as
\begin{equation}
\widetilde{n}_{k}= \int_{-\infty}^{\widetilde{\lambda}_{k}} dx
\zeta (x)+\sum_{\mu=1}^{M}a_{2\mu}\zeta_{2\mu-1}
(\widetilde{\lambda}_{k}) ,
\label{strut}
\end{equation}
where $\zeta (x)$ is the averaging function chosen as
$$
\zeta (x)={1 \over \sqrt{\pi}} e^{-\displaystyle x^{2}}.
$$
The second term in Eq. (\ref{strut}) contains the so-called
Strutinsky curvature corrections and is given by
$$
\zeta_{n}(x)={(-1)^{n} \over \sqrt{\pi}} e^{-\displaystyle 
 x^{2}}H_{n}(x) ,
$$ 
where $H_{n}(x)$ are the Hermite polynomials, $a_{2n}=(-1)^{n}/(2^{2n}n!)$, 
and $\widetilde{\lambda}_{k}=(E_{F}-E_{k})/\gamma$. The quantities
$E_{F}$ and $E_{k}$ are the Fermi energy and single-particle energies
in the mean field $V_{eq}(r)$, respectively, and $\gamma$ is an averaging
parameter.

The results of numerical calculations (see also Ref. \cite{PrShKo})
of the smooth distribution function $\widetilde{f}_{eq}(\epsilon)$
of Eq. (\ref{fw}) are shown in Fig. 1. The solid line 1 gives 
the behaviour of the smooth distribution function for the value of the
smearing parameter  $\gamma=2.5\cdot \hbar \,\Omega$ and $2M=6$, the order 
of the curvature correction polynomial in Eq. (\ref{strut}).
We point out that these values of the smearing parameter and correction 
polynomial are localized  in the so-called plateau region for the shell 
correction $\delta U$ to the binding energy, i.e., where $\delta U$ does
not depend on $\gamma$, Ref. \cite{KoKoStKhv}.

The smooth distribution function $\widetilde{f}_{eq}(\epsilon)$
exhibits oscillations caused by particle reflections on the
potential surface. The mean behaviour (i.e., without the
oscillations) of $\widetilde{f}_{eq}(\epsilon)$ can be approximated
by the Fermi function $f_{eq,F}(\vec{r}, \vec{p})$ of Eq. (\ref{ff}),
where the chemical potential $\lambda$ is fixed by the condition of 
conservation of particle number A:
$$
A=g\int {d\vec{r}d\vec{p} \over (2\pi \hbar)^{3}}
f_{eq,F}(\vec{r}, \vec{p}).
$$ 
The solid line 2 in Fig. 1 shows the behaviour of the Fermi distribution
function 
$f_{eq,F}(\vec{r}, \vec{p})$ of Eq. (\ref{ff}) with the parameter $a$ 
taken from Eq. (\ref{aapx}). The dashed line in Fig. 1 gives the
simple Thomas-Fermi approximation Eq. (\ref{ftf}).

In Fig. 2 we show the smooth distribution functions
$\widetilde{f}_{eq}$  as functions of the dimensionless kinetic
energy $\epsilon_{kin} = p^{2}/m\hbar\Omega$ for different
distances $r$. We can see from this figure that the
diffusivity parameter $a$ is almost independent of the distance
$r$. This fact is a feature of the HO potential well.
Note that the analogous diffusivity parameter for the Woods-Saxon
potential is a strongly $r$-dependent function increasing near
the potential wall, see below and Ref. \cite{DuRaSc,KoPl377}. 
Following Ref. \cite{KoPl377},
the diffusivity parameter $a$ of the quantum distribution function
$f_{eq}$ in momentum space can be estimated using the expansion
of $f_{eq}$ in Hermite polynomials. The result reads 
\begin{equation}
a \simeq \sqrt{G_{2} + \left[{G_{3} \over 2} \right]^{2/3}}.
\label{a}
\end{equation}
The parameters $G_2$ and $G_3$ depend on the mean field
$V(r)$ and are given by (in the lowest order of $\hbar^2$)
\begin{equation}
G_{2} =-{\hbar^{2}\over 4m}\left[{2\over r}V_{eq}^{\prime}(r) +
V_{eq}^{\prime\prime}(r)\right], \ \ \
G_{3}=-{\hbar^{2}\over 4m}
\left[(V_{eq}^{\prime}(r))^{2}+{p^{2} \over 3m}
\left[{2\over r} V_{eq}^{\prime}(r) + V_{eq}^{\prime\prime}(r) \right]\right],
\label{GG}
\end{equation}
where $prime$ means an $r$ derivative.
For the HO potential one has $G_{2}=-3(\hbar\Omega)^{2}/4$ 
and $G_{3} \simeq -(\hbar\Omega)^{3}\lambda/2 $.
The diffusivity parameter $a$ can be also estimated from a
fit of the smooth distribution function $\widetilde{f}_{eq}$
and its derivative $d\widetilde{f}_{eq} / dE$ to the corresponding
values of the Fermi distribution function Eq. (\ref{ff}) within some
smeared out interval $\widetilde{\gamma} \leq \gamma$ 
near the Fermi energy $E_F$. Namely, one has the following estimate:
\begin{equation}
a=\displaystyle {1 \over 
\widetilde{\gamma}}  \int^{E_{F}+\widetilde{\gamma} /2}_{E_{F}-
\widetilde{\gamma}/2} dE
{\widetilde{f}_{eq}(\widetilde{f}_{eq}-1) \over d\widetilde{f}_{eq} / dE}.
\label{adif}
\end{equation}
Note that Eq. (\ref{adif}) gives an exact result for the diffusivity
parameter $a$ if the distribution function $\widetilde{f}_{eq}$
coincides with the Fermi distribution function of Eq. (\ref{ff}).

The solid lines 1 and 2 in Fig. 3 show the numerical results for
the parameter $a$ as a function of mass number $A$ obtained using
both expressions (\ref{a}) and (\ref{adif}), respectively.
A discrepancy between the results presented by curves 1 and 2
appears because the expression (\ref{a}) takes into account
the lowest orders of expansion of the distribution
function $\widetilde{f}_{eq}$ over the Hermite polynomials;
see Ref. \cite{KoPl377}. We have also established the following
simple $A$ dependence for the diffusivity parameter $a$ for
the spherical HO potential well, obtained from Eq. (\ref{adif})
with $\widetilde{\gamma} = \gamma/2 = 1.25\,\hbar \Omega$,
\begin{equation}
a \approx 10.2 A^{-1/6} \,MeV. 
\label{aapx}
\end{equation} 
The result of the numerical calculation of the diffusivity parameter
$a$  Eq. (\ref{aapx}) is shown in Fig. 3 by the dashed line.

We will return now to the problem of evaluating 
the relaxation time, Eq. (\ref{tau2}). 
In Fig. 4 we have plotted the ratio $q$ as obtained from Eq. (\ref{q})
as a function of the mass number $A$ (curve HO for the harmonic
oscillator potential (\ref{HO})). We have used here the Fermi
distribution function Eq. (\ref{ff}) with parameter $a$ from Eq.
(\ref{aapx}). As can be seen from Fig. 4, the effect of the
quantum oscillations for the smeared-out distribution function
$f_{eq} = \widetilde{f}_{eq}$ leads to an essential compensation
of the gain and loss probability fluxes. A small nonzero contribution
(about 10 - 20 \% in Fig. 4) remains in part because of the 
Abrikosov-Khalatnikov transformation Eq. (\ref{ach}) used earlier, 
which implies a localization of the momentum $\vec{p}_j$ near the 
Fermi surface in Eq. (\ref{dst}). In order to check this 
statement, we will use more general transformation of the momentum 
integrals in the collision integral, Eq. (\ref{dst}), with an arbitrary 
value of the momentum $p_j$, suggested in Ref. \cite{SaYo}:
\begin{eqnarray} 
&\displaystyle \int {d\vec{p}_{2} d\vec{p}_{3} d\vec{p}_{4}
\over (2 \pi \hbar)^{6}} \delta (\Delta \vec{p})\delta (\Delta E) (...)&
\nonumber \\
&= \displaystyle {m^3 \over 16 \pi^{4} \hbar^{4}\,p_1} \int dE_{2} dE_{3}
\left[ \sqrt{p_{1}^{2} + p_{2}^{2} + 2 p_{1} p_{2} \kappa } -
\sqrt{p_{1}^{2} +p _{2}^{2} - 2 p_{1} p_{2} \kappa } \right ] (...) ,
&
\label{sayo}
\end{eqnarray}
where $\kappa = {\rm min} \left (1,\sqrt{p_{3}^{2} (p_{1}^{2} + 
p_{2}^{2} - p_{3}^{2} ) / p_{1}^{2} p_{2}^{2}} \right) $.
The solid line 1 in Fig. 5 gives the behaviour of the ratio $q$ as
a function of the smoothing parameter $\gamma$ as obtained from
Eq. (\ref{q}) by applying the transformation Eq. (\ref{sayo}) to the
calculations of the probability fluxes, Eqs. (\ref{gain}) and
(\ref{loss}). A strong deviation of this result for $q$ from
the analogous one obtained using the Abrikosov-Khalatnikov procedure,
Eq. (\ref{ach}), (the solid line 2) appears for small magnitudes
of smoothing parameter $\gamma$. This is because the smeared-out 
distribution function $\widetilde{f}_{eq}$ oscillations
grow with the decrease of the smoothing parameter $\gamma$ and 
the accuracy of the Abrikosov-Khalatnikov transformation Eq. 
(\ref{ach}) also decreases.

The results of the numerical calculations of the local relaxation time 
(more precisely, the inverse value $\hbar /\tau_{2}(r,\omega_{0})$) of 
Eq. (\ref{tau22}) for the isoscalar  GQR in the
nucleus with $A = 224$ are shown in Fig. 6. We point out that the use
of the Fermi distribution function Eq. (\ref{ff}) instead of
$\widetilde{f}_{eq}$ in Eq. (\ref{tau22}) (the solid curve 2
in Fig. 6) leads to much stronger damping than the analogous
calculations with $\widetilde{f}_{eq}$  from Eq. (\ref{fw})
(solid curve 1).

To give a simple phenomenological prescription for the removal of
the nonphysical probability fluxes (\ref{gain}) and (\ref{loss})
in the ground state of a system, we will introduce the
modified distribution function 
\begin{equation}
{\cal F}_{eq}=f_{eq,F} + \eta \Delta \widetilde{f}_{eq}.
\label{ffs}
\end{equation}
Here, $f_{eq,F}$ is the Fermi distribution function Eq. (\ref{ff})
with diffusivity parameter $a$ from Eq. (\ref{aapx}), and  
$\Delta \widetilde{f}_{eq}$ is given by
$$
\Delta \widetilde{f}_{eq}=\widetilde{f}_{eq}-f_{eq,F} ,
$$   
where $\widetilde{f}_{eq}$ is the smooth distribution function
of Eq. (\ref{fw}). A numerical calculation of the probability fluxes
$J_{eq, gain}$, Eq. (\ref{gain}), and  $J_{eq, loss}$, Eq. (\ref{loss}), 
with  $\widetilde{f}_{eq}$ replaced by ${\cal F}_{eq}$ shows
that both probability fluxes disappear at $\eta = 0.86$ for the
nucleus with $A$ = 224. The corresponding distribution function
${\cal F}_{eq}$ is shown in Fig. 1 as the dotted line.
The local relaxation time from Eq. (\ref{tau22}) obtained with
modified distribution function from Eq. (\ref{ffs}) is shown in Fig. 6 
as solid curve 3.  It should be remembered that the  two last 
results (curves 1 and 3) are more correct in the sense of the
compensation of the probability fluxes in the ground state of the
system. Both these results show  strong oscillations
of the local relaxation time within the nuclear volume and  
amplification of the damping in the surface region of the nucleus. 
This feature of the finite Fermi system arises due to the fact that
the equilibrium distribution function $\widetilde{f}_{eq}$,
Eq. (\ref{fw}) fluctuates  strongly and contains the diffusivity 
of the Fermi surface. It is interesting to note that, excluding
the oscillations in the nuclear volume, the result noted as 
curve 3 in Fig. 6 with full compensation of the probability
fluxes in the ground state is in good agreement with the one
$(\hbar/\tau_{2})_{TF}$
obtained using the simple Thomas-Fermi distribution function of
Eq. (\ref{ftf}) (the dashed line in Fig. 6 with $(\hbar/\tau_{2})_{TF}
=3(\hbar\omega_{0})^{2}/ (4\pi^{2}\alpha)$).

We point out that the quantum calculations represented in Fig. 6 by 
curves 1 and 3 have been done with a quite large smoothing parameter 
$\gamma = 2.5 \cdot \hbar\Omega$. A decrease of $\gamma$ leads to negative 
values of the local relaxation time $\tau_{2}(r,\omega_{0})$ 
in some regions  of $r$. The behaviour of the damping
factor $\xi (r,\omega_{0})$ is shown in Fig. 7. It is necessary to
stress that nonzero damping in a cold Fermi system appears only
because of the memory effects in the collision integral (\ref{dst}).

The results of numerical calculations of the width $\Gamma$ of the 
isoscalar GQR as a function of the mass number $A$ are shown in Fig. 8.
As can be seen from this figure, the smooth distribution
function $\widetilde{f}_{eq}$ from Eq. (\ref{fw}) leads to
the contribution of the collisional relaxation to the isoscalar 
GQR width (solid line 1) which does not exceed 30-50\% of the 
experimental values. This result is similar to the one obtained
with the sharp Thomas-Fermi distribution function Eq. (\ref{ftf})
(the dashed line) and agrees with the earlier calculation of the internal 
collisional width of the isoscalar GQR \cite{KoPlSh1}. In contrast, the
analogous calculation with the smooth Fermi distribution function (\ref{ff}) 
(solid line 2) largely overestimates the contribution 
of collisional damping. Note that the small nonzero probability
fluxes appearing in $J_{eq, gain}$ (\ref{gain}) and  $J_{eq, loss}$
(\ref{loss}) evaluated with $\widetilde{f}_{eq}$ from Eq. (\ref{fw})
lead to a very small contribution to the  final result
for the width. This can be seen in Fig. 8 from a comparison of both
curves 1 and 3.

\medskip
{\bf B. Woods-Saxon potential}
\medskip

We have applied above the HO Wigner distribution function to the
calculations of the collision integral in a finite Fermi system.
This distribution function contains both important ingredients
of influence of the multiple particle reflections from the potential
surface:  the diffusivity and the oscillations of the distribution
function in momentum space. However, the realistic nuclear potential well 
has a finite depth, providing  stronger surface effects than the ones
in the HO mean field. We will give below an analysis for the case
of the WS potential in the form
\begin{equation}
V_{WS}(r)=V_{0}/(1+{\rm exp}[(r-R_{0})/ d]).
\label{WS}
\end{equation}
We have adopted the following parameters $V_0 =-44\,MeV\,\,
R_0 =1.27\,A^{1/3}\,\,$ and $d = 0.67\,fm$.

An exact quantum calculation of the equilibrium distribution
function $f_{eq}(\vec{r}, \vec{p})$ for the WS potential is a rather 
complicated problem. We will use the result of Ref. 
\cite{KoPl377} for the semiclassical expansion of 
$f_{eq}(\vec{r}, \vec{p})$ in the Hermite polynomials. This gives
\begin{equation}
f_{eq}(\vec{r},\vec{p}) = \sum_{n=0}^{N} {b_{n} 
(-1)^n \over n!} \Phi^{(n)} \left( {x-\mu \over \sigma} \right) 
\Big\vert_{x=E_{F}}.
\label{distrib.funk.for.ws}
\end{equation}
Here $\Phi(y)=[1+{\rm erf}(y/\sqrt{2})]/2$ is the normal distribution
function, ${\rm erf}(y)$ is the error function, $\Phi^{(0)} \equiv \Phi$ 
and the index $n$ at $\Phi^{(n)}((x- \mu )/\sigma)$ denotes the 
$x$ derivative of $n$ order. We will take into account terms with 
$n \leq 3$ in the expansion (\ref{distrib.funk.for.ws}). The corresponding 
coefficients $b_n$ are given by
$$
b_{0}=1; \ b_{1}=E(\vec{r},\vec{p})-\mu; \ 
b_{2}=G_{2}+b_{1}^{2}-\sigma^{2}; \ 
b_{3}=G_{3}+b_{1}^{3}+3b_{1}(G_{2}-\sigma^{2}),
$$
where $G_{2}$ and $G_{3}$ are obtained from (\ref{GG}). The coefficients
$\mu$ and $\sigma$ can be found by solving the following
nonlinear equations:
$$
b_{4} \equiv b_{1}^{4} + 6 b_{1}^{2} ({G}_{2} - \sigma^{2}) 
+ 4 b_{1} {G}_{3} + 
3 \sigma^{2} (\sigma^{2} - 2 {G}_{2}) = 0,
$$ 
$$
b_{5} \equiv 15 \sigma^{4} b_{1} - 
10 \sigma^{2} (b_{1}^{3} + 3 b_{1} {G}_{2} + {G}_{3}) 
- b_{1}^{5} - 10 b_{1}^{3} {G}_{2} -10 b_{1}^{2} {G}_{3} 
= 0.
$$

Following Eq. (\ref{ff}), the mean behaviour  
of $f_{eq}(\vec{r},\vec{p})$, Eq. (\ref{distrib.funk.for.ws}), 
can be approximated by the following Fermi function:
\begin{equation}
f_{eq,WS}(\vec{r},\vec{p})=\left(1+\exp\left[{E(\vec{r},\vec{p})- \lambda
 \over a(r)}\right]\right)^{-1}.
\label{ffws}
\end{equation}
Here $E(\vec{r},\vec{p})$ is the classical energy of the
particle in the WS potential
$$
E(\vec{r},\vec{p}) = {p^2\over 2m} + V_{WS}(r).
$$
The local diffusivity $a(r)$ is evaluated from Eq.
(\ref{GG}) with ${G}_{j}$ taken in the so-called surface approximation 
($V_{eq}^{\prime \prime}(r) \approx 
V_{eq}^{\prime \prime}(R_0) =0$) \cite{KoLuPl}
$$
G_{2}=-{\hbar^{2} \over 4m}\left[{2 \over R_0} 
V_{eq}^{\prime}(r) \right ]; 
\ \ \ \ 
G_{3}=-{\hbar^{2} \over 4m}\left[(V_{eq}^{\prime})^{2} 
+{p_{F}^{2} \over 3m}\left[{2 \over R_0} 
V_{eq}^{\prime}(r)\right ] \right ],
$$
where $p_{F}(r)= \hbar (3 \pi^{2} \rho_{eq}(r) / 2)^{1/3}$ is 
the Fermi momentum taken in the local density approximation.
The results of numerical calculations of the parameter $a(r)$ are
plotted in Fig. 9 for both HO and WS distribution functions.
We can see that the parameter $a(r)$ for the WS potential well
is peaked in the vicinity of the nuclear surface.
The distribution function $f_{eq}(\vec{r}, \vec{p})$ smeared out
over angles in $\vec{p}$ space is plotted in Fig. 10. It contains
both diffusivity and oscillations in $\vec{p}$ space which
depend on the distance $r$. In Fig. 4 we show the ratio $q$ as 
defined by Eq. (\ref{q}) (curve WS).
We have used there the Fermi distribution function, Eq. (\ref{ffws}), 
instead of the smooth distribution function $f_{eq,F}$ in 
Eq. (\ref{q}). One can see that the occurrence of the
quantum oscillations in the smeared-out distribution function
$\widetilde{f}_{eq}$ leads to an essential compensation
of the gain and loss probability fluxes.

The local relaxation parameter  (\ref{tau22}) and the damping factor
(\ref{xi}) are shown in Figs. 11 and 12, respectively. It can be seen 
that the collisional damping is more pronounced in the case of the
WS potential than in the HO one shown in Figs. 6 and 7.
We point out also that the occurrence of the quantum oscillations
in the equilibrium distribution function  strongly reduces the
relaxation processes in the finite Fermi system.
The total collisional width $\Gamma$ of the 
isoscalar GQR as a function of the mass number $A$ evaluated for
WS potential is shown in Fig. 13. The final result (solid
line 1) agrees with above calculation of the collisional width
in the HO potential; see the dashed line in Fig. 8.

\section{Summary and Conclusions}

Starting from the collisional kinetic equation we have derived
the isoscalar GQR width for a finite nucleus taking into account the
memory effects and the peculiarities of the equilibrium
distribution function $f_{eq}$ caused by the multiple reflections of
particles on the potential wall. The equilibrium distribution
function contains both smooth and the oscillating components.
The smooth component of $f_{eq}$ can be approximated by the Fermi 
distribution (\ref{ff}) with an $r$-dependent diffusivity parameter $a(r)$ 
in momentum space. The diffusivity parameter $a(r)$ is almost independent of 
the distance $r$ in the case of the harmonic oscillator potential well 
and is a strongly $r$-dependent function in the case of the Woods-Saxon 
potential, increasing near the potential wall. We have demonstrated
numerically that the smooth part of the equilibrium distribution function 
can be satisfactorily described by using the semiclassical expansion of 
$f_{eq}(\vec{r}, \vec{p})$ on the Hermite polynomials; see Ref.
\cite{KoPl377}.

It was shown that the general
condition for the disappearance of the gain and loss probability 
fluxes in the ground state of system can be reached due to the occurrence
of quantum oscillations in the distribution function in momentum
space in the nuclear volume. The diffuse tail of the distribution
function in momentum space leads to an increase of the collisional 
damping of the collective motion in the surface region of the nucleus
and thus to an increase of the isoscalar GQR width. However this increase is 
strongly reduced  due to the above-mentioned oscillations of 
the equilibrium distribution functions appearing in the collision
integral Eq. (\ref{dst}). As a result the collisional width of the
isoscalar GQR does not exceed 30-50\% of the experimental value
and agrees with the estimates of the width where the sharp
Thomas-Fermi distribution function  (i.e., in the absence of the
diffusivity and quantum oscillations) is used. The collisional
damping in a cold Fermi system arises only because of
memory effects in the collision integral (\ref{dst}). 

To describe the experimental values of the multipole giant resonances 
additional contributions from other spreading sources, such as the
fragmentation width in random phase approximation calculations or its 
representation through the one-body dissipation (see Ref. \cite{KoPlSh2})
have to be taken into account. We pointed out also that 
surface effects in the collisional damping are manifested
more distinctly in the Woods-Saxon potential where the diffusivity
parameter $a(r)$ of the distribution function in momentum space
is $r$ dependent and increases within the surface region of the nucleus.

\newpage

{\bf ACKNOWLEDGEMENTS}

This work was supported in part  by  the U.S. National Science 
Foundation under Grant No. PHY-9413872 and the INTAS under Grant No. 93-0151.
We are grateful for this financial support. One of us (V.M.K.) thanks
Professor P.Schuck for stimulating discussion and
the Cyclotron Institute at Texas A\&M University for the kind
hospitality.

\newpage

\begin{figure}[t]
\begin{center}
\leavevmode
\epsfverbosetrue
\epsffile{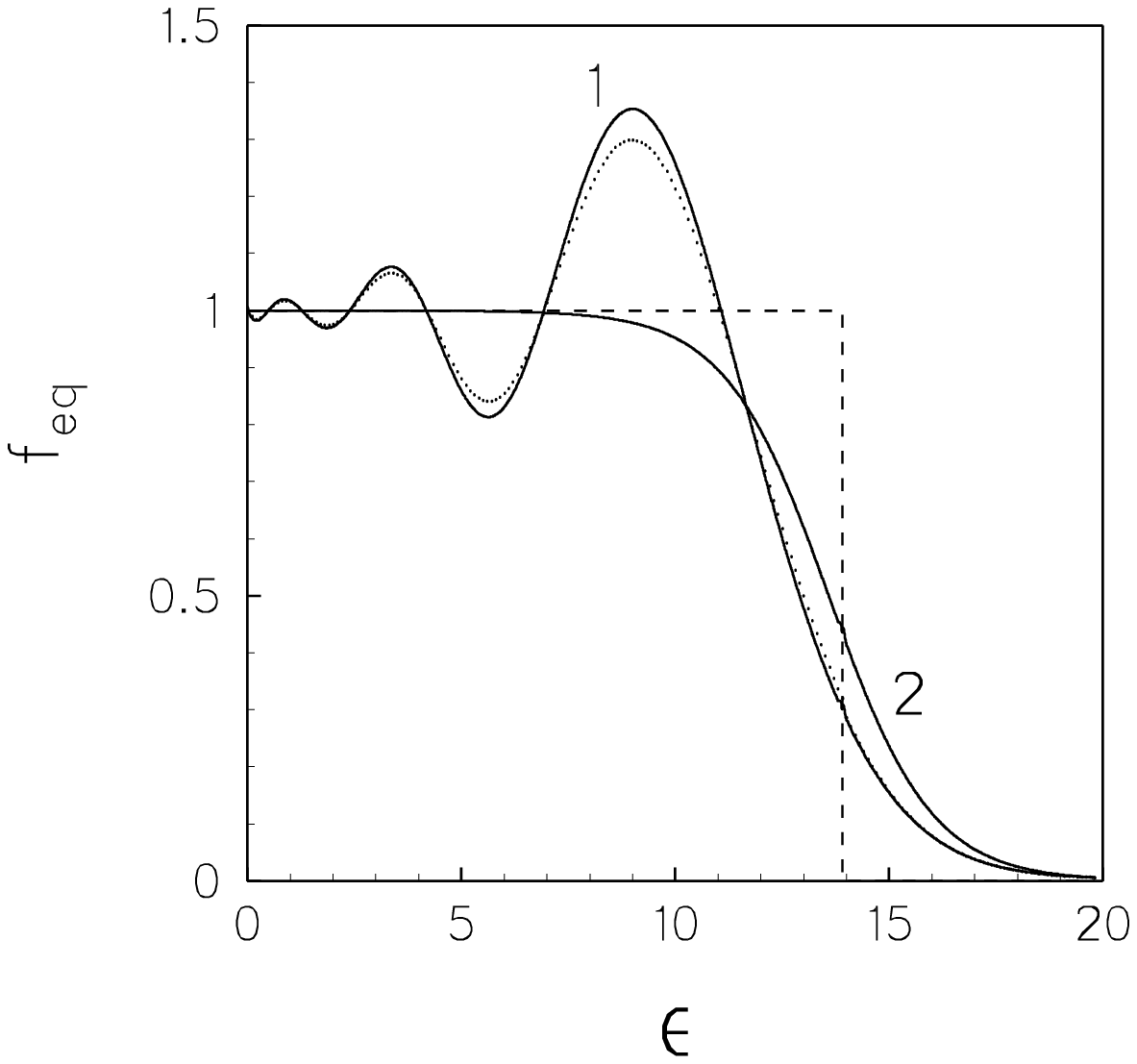}
\end{center}
\caption{The equilibrium distribution function 
as a function of the dimensionless parameter $\epsilon$, for a
nucleus with mass number $A$= 224, calculated for a spherical HO
potential. Solid lines 1 and 2 show the smooth distribution
function obtained using an averaging procedure of Eq. (50)
with $\gamma = 2.5\cdot\hbar\Omega$ and the Fermi distribution
function of Eq. (2) with parameter $a$ from Eq. (55),
respectively. The dashed line shows the Thomas-Fermi distribution
function (1). The dotted line shows the distribution function 
of Eq. (57) with $\eta = 0.86$ providing the disappearance
of the probability fluxes in the ground state of the nucleus.}
\end{figure}

\newpage

\begin{figure}[t]
\begin{center}
\leavevmode
\epsfverbosetrue
\epsffile{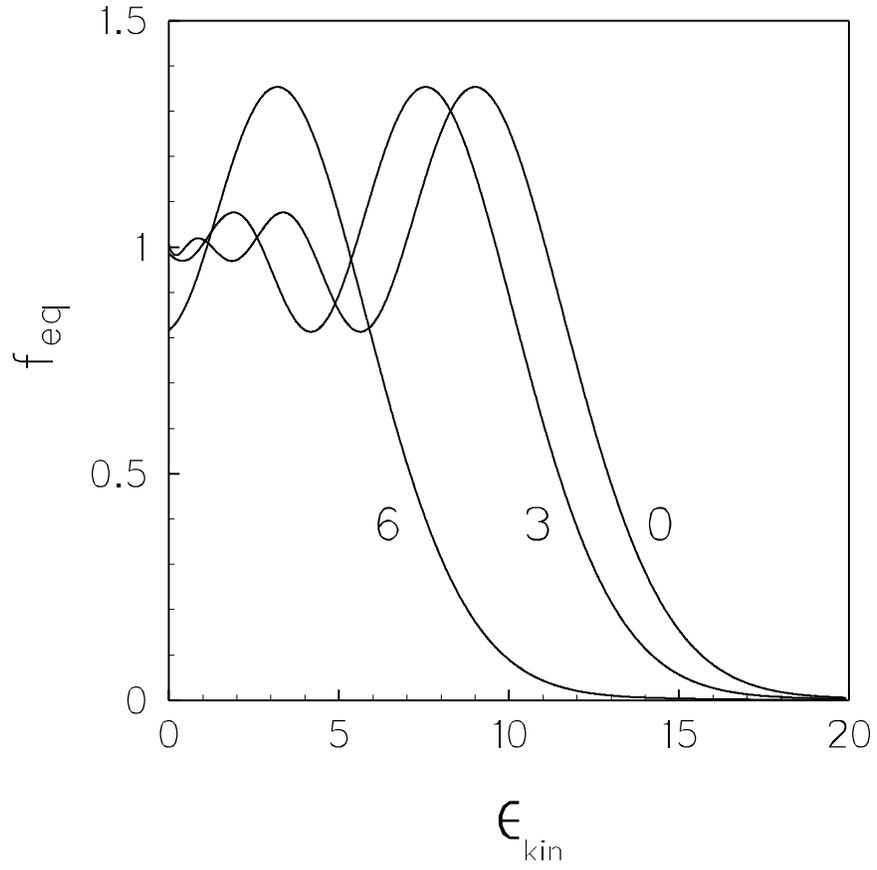}
\end{center}
\caption{The smooth distribution function
of Eq. (50) as a function of the dimensionless kinetic
energy $\epsilon_{kin}=p^{2}/m\hbar\Omega$, 
for a nucleus with mass number $A$= 224, calculated for a spherical HO
potential with $\gamma = 2.5 \cdot \hbar \Omega$. The different curves 
correspond to the distances $r = 0,\, 3,$ and $6\,fm$ to the center
of the nucleus.}
\end{figure}

\newpage

\begin{figure}[t]
\begin{center}
\leavevmode
\epsfverbosetrue
\epsffile{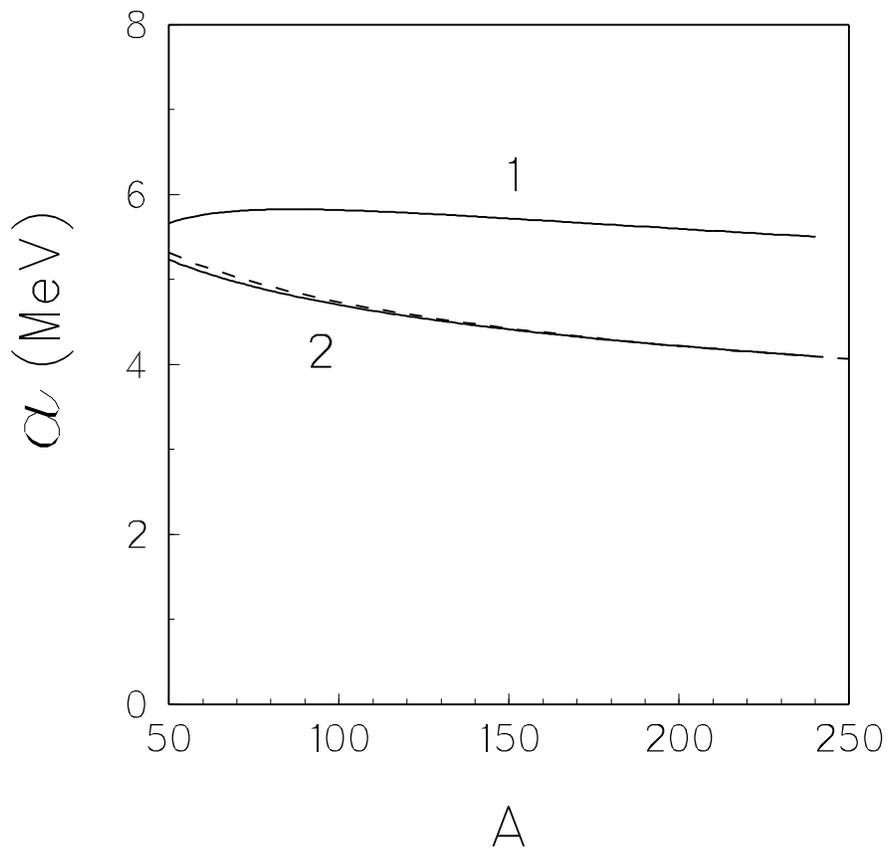}
\end{center}
\caption{The diffusivity parameter of the smooth distribution function
of Eq. (50) versus the mass number $A$, calculated for a spherical HO 
potential with $\gamma = 2.5 \cdot \hbar \Omega$: For curve 1 we use
Eq. (52), for curve 2 we use Eq. (54) and the dashed
curve is obtained from the fitting formula of Eq. (55).}
\end{figure}

\newpage

\begin{figure}[t]
\begin{center}
\leavevmode
\epsfverbosetrue
\epsffile{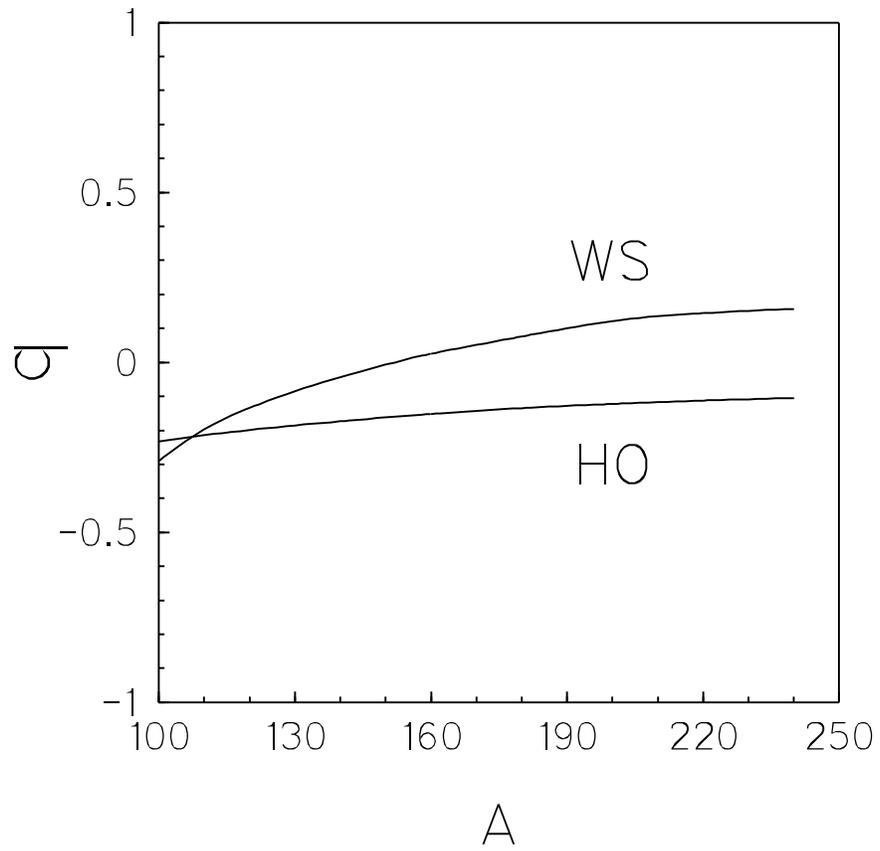}
\end{center}
\caption{The relative probability fluxes as given by Eq. (48) versus 
the mass number $A$ for both HO and WS potentials.}
\end{figure}

\newpage

\begin{figure}[t]
\begin{center}
\leavevmode
\epsfverbosetrue
\epsffile{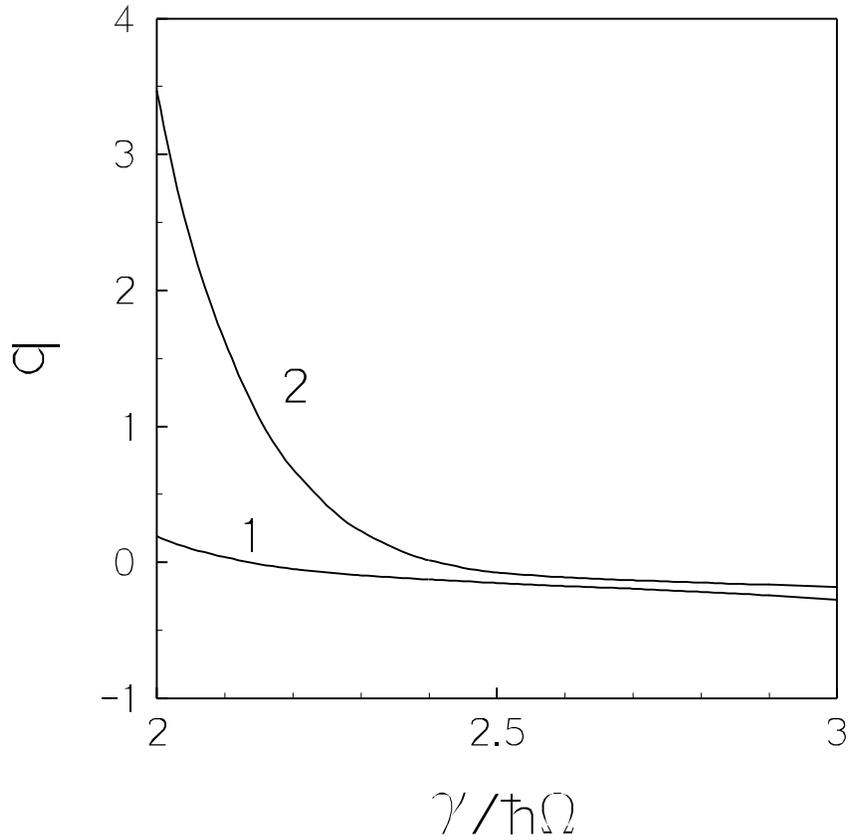}
\end{center}
\caption{Dependence of the relative probability fluxes $q$ on the averaging 
parameter $\gamma$ in units of $\hbar\Omega$: For solid line 2 we
use the Abrikosov-Khalatnikov transformation Eq. (13), and
for solid line 1 we use the transformation Eq. (56), for the
momentum integrals in the collision integral Eq. (9).
The calculations were performed for the nucleus with
$A$=224 in the HO potential well.}
\end{figure}

\newpage

\begin{figure}[t]
\begin{center}
\leavevmode
\epsfverbosetrue
\epsffile{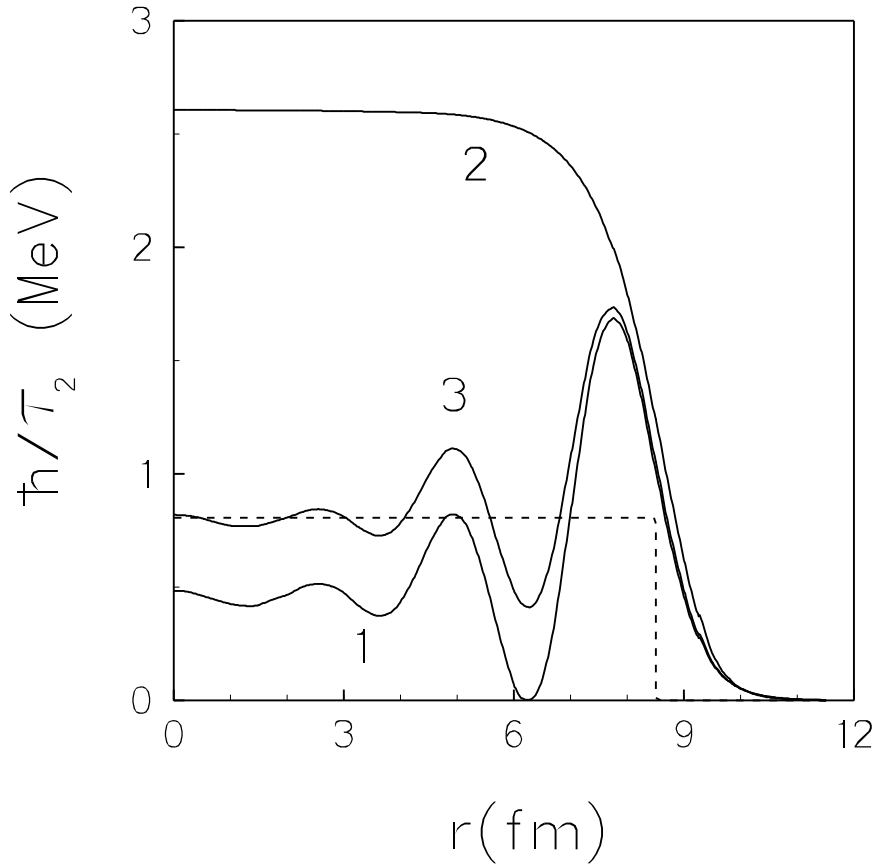}
\end{center}
\caption{The local damping parameter $\hbar/\tau_{2}(r,\omega_{0})$
for the nucleus with $A =224$, calculated for a spherical HO
potential. The different curves correspond to the different
equilibrium distribution function in Eq. (42): 
For curve 1 we use the smooth distribution function of
Eq. (50) with averaging parameter $\gamma = 2.5 \cdot \hbar \Omega$, 
for curve 2 we use the Fermi distribution function (2) 
with $a$ from Eq. (55), and for curve 3 we use the distribution
function from Eq. (57). The dashed line shows the result
obtained using the Thomas-Fermi distribution function of
Eq. (1).}
\end{figure} 

\newpage

\begin{figure}[t]
\begin{center}
\leavevmode
\epsfverbosetrue
\epsffile{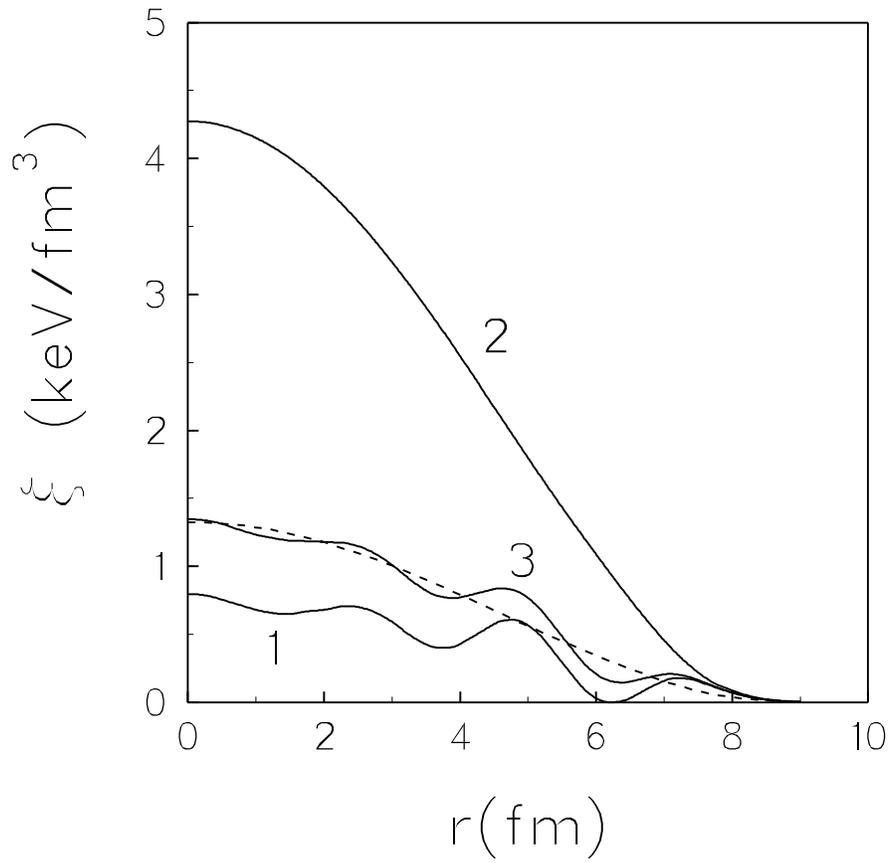}
\end{center}
\caption{The damping factor $\xi (r,\omega_{0})$ as a function of the 
distance $r$. The notations is the same as in Fig. 6.}
\end{figure}

\newpage

\begin{figure}[t]
\begin{center}
\leavevmode
\epsfverbosetrue
\epsffile{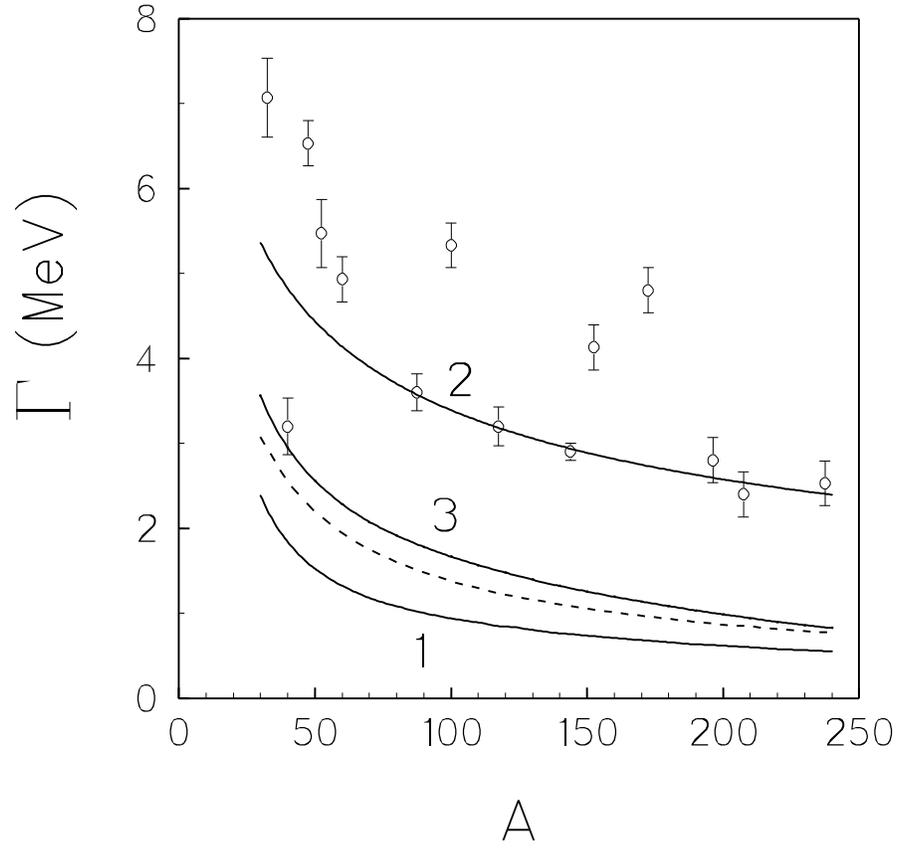}
\end{center}
\caption{The collisional width of the isoscalar giant quadrupole resonances
obtained by using Eq. (40) with $\xi (r,\omega_{0})$
from Fig. 7.  The notations is the same as in Fig. 7. 
The experimental data were taken from Refs. [31, 32].}
\end{figure}

\newpage

\begin{figure}[t]
\begin{center}
\leavevmode
\epsfverbosetrue
\epsffile{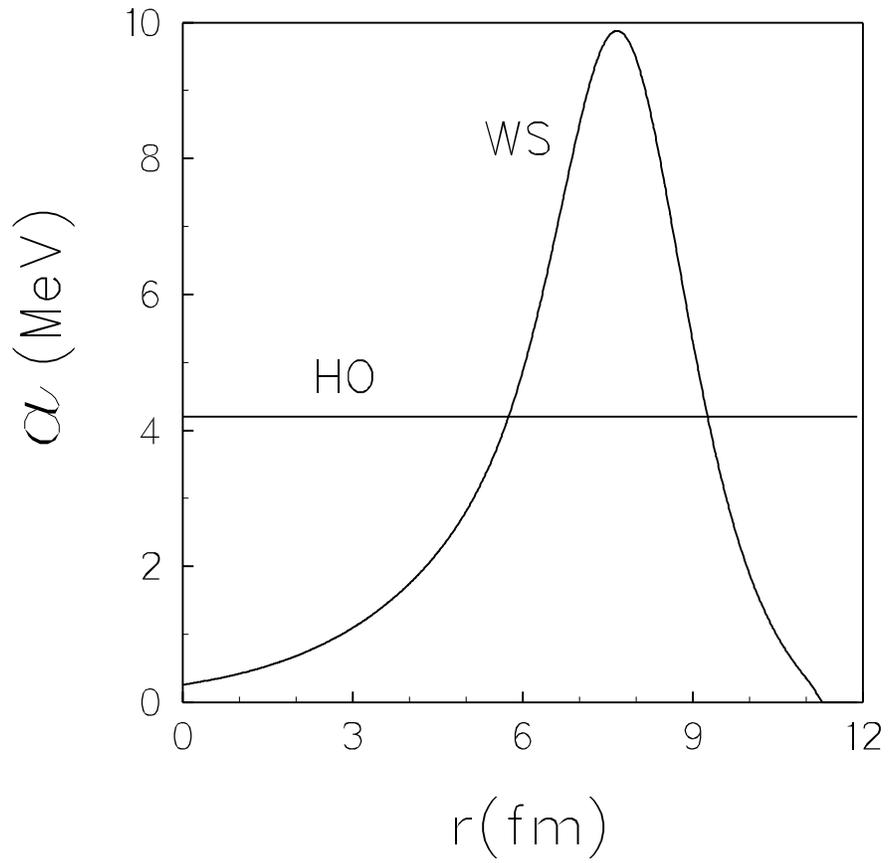}
\end{center}
\caption{The diffusivity parameter $a$ of the equilibrium
distribution function in momentum space as a function of the
distance $r$ to the center of the nucleus for both harmonic oscillator
(curve HO) and Woods-Saxon (curve WS) potentials.}
\end{figure}

\newpage
\begin{figure}[b]
\begin{center}
\leavevmode
\epsfverbosetrue
\epsffile{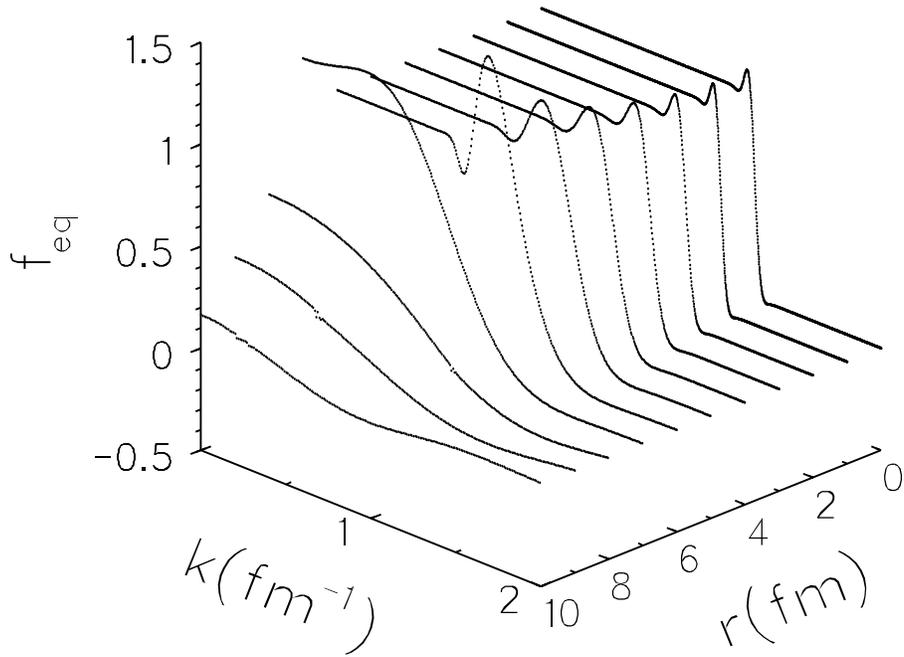}
\end{center}
\caption{The equilibrium distribution function in
the WS potential as obtained from the semiclassical expansion on
the Hermite polynomials Eq. (59), with N = 3.}
\end{figure}

\newpage

\begin{figure}[t]
\begin{center}
\leavevmode
\epsfverbosetrue
\epsffile{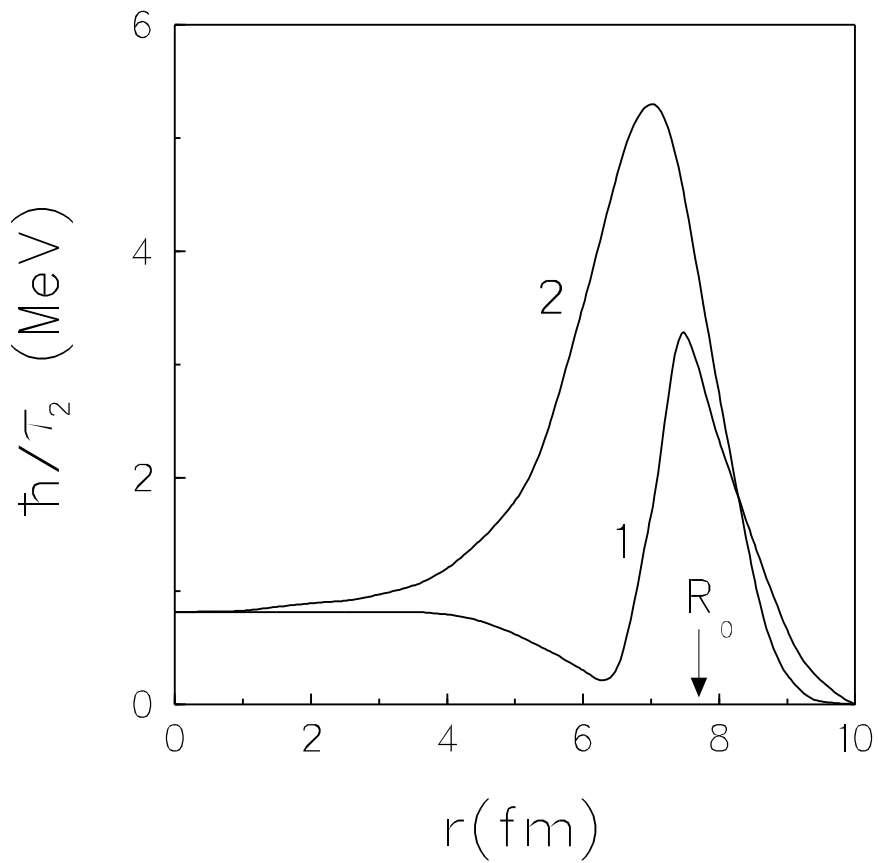}
\end{center}
\caption{The local damping parameter 
$\hbar/\tau_{2}(r,\omega_{0})$ for the nucleus with $A=224$, 
calculated for a spherical WS potential. For curve 1 we use
the equilibrium distribution function (59)
and for curve 2 we use the Fermi distribution function (60).}
\end{figure}

\newpage

\begin{figure}[t]
\begin{center}
\leavevmode
\epsfverbosetrue
\epsffile{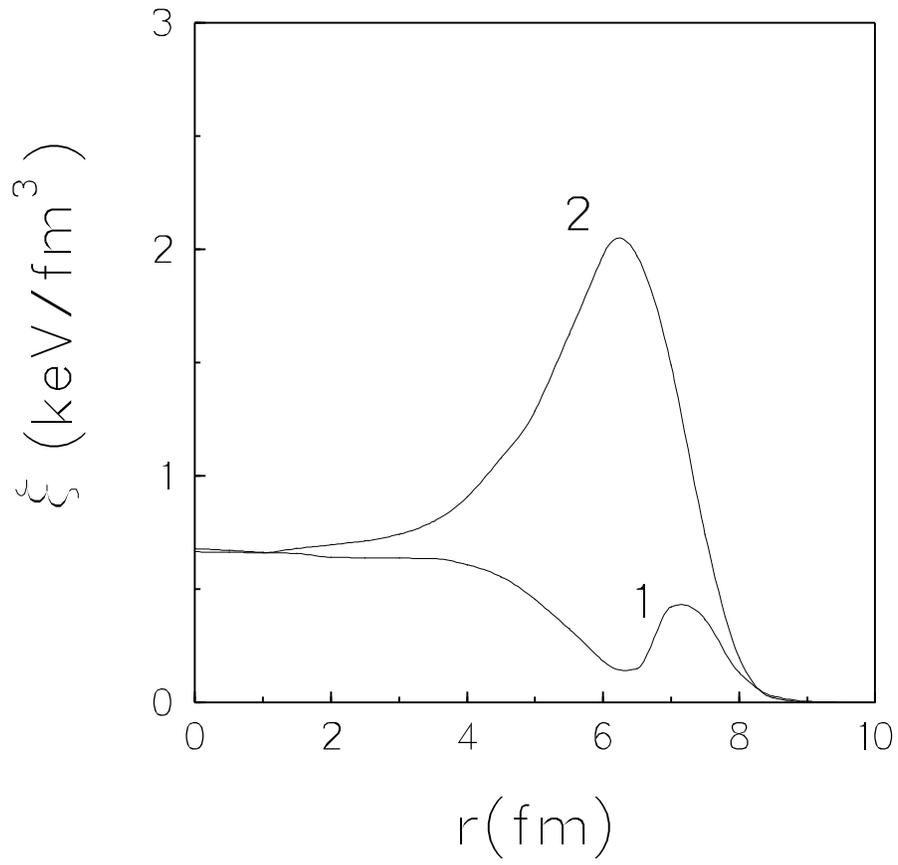}
\end{center}
\caption{The damping factor $\xi$ in the
case of the WS potential. The notations is the same as in Fig. 11.}
\end{figure}

\newpage

\begin{figure}[t]
\begin{center}
\leavevmode
\epsfverbosetrue
\epsffile{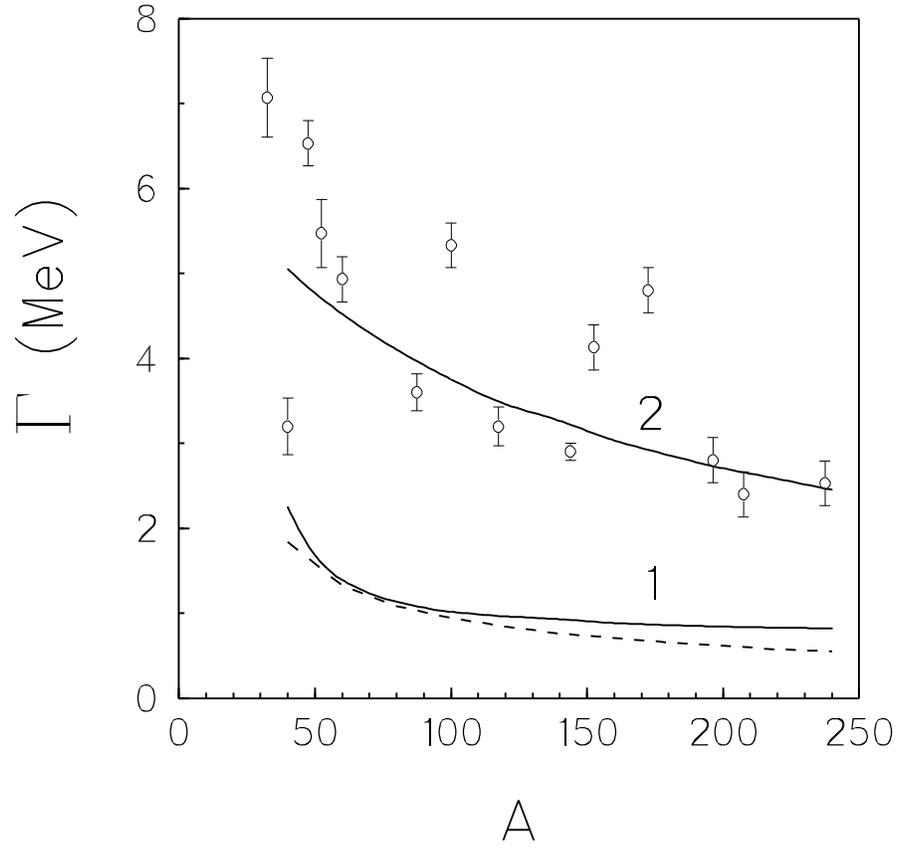}
\end{center}
\caption{The collisional width of the isoscalar giant 
quadrupole resonances obtained by using Eq. (40) with $\xi 
(r,\omega_{0})$ from Fig. 12.  The notations of curves 1 and 2 is
the same as in Fig. 12. The dashed curve is from Fig. 8.
The experimental data were taken from Refs. [31, 32].}
\end{figure}

\end{document}